\documentclass[article,nofootinbib,notitlepage]{revtex4-1}
\usepackage{amsmath}
\usepackage{graphicx}
\usepackage{dcolumn}
\usepackage{bm}
\usepackage[latin1]{inputenc}
\usepackage{amsthm,latexsym,amssymb,amsfonts,array}

\begin{document}

\title{Relational Mechanics as a gauge theory}
\author{Rafael Ferraro\bigskip}
\email{ferraro@iafe.uba.ar}
\thanks{member of Carrera del Investigador Cient\'{\i}fico (CONICET, Argentina). }
\affiliation{Instituto de Astronom\'\i a y F\'\i sica del Espacio
(IAFE, CONICET-UBA), Casilla de Correo 67, Sucursal 28, 1428 Buenos
Aires, Argentina.} \affiliation{Departamento de F\'\i sica, Facultad
de Ciencias Exactas y Naturales, Universidad de Buenos Aires, Ciudad
Universitaria, Pabell\'on I, 1428 Buenos Aires, Argentina.}

\begin{abstract}
\textit{Absolute space} is eliminated from the body of mechanics by
gauging translations and rotations in the Lagrangian of a classical
system. The procedure implies the addition of compensating terms to
the kinetic energy, in such a way that the resulting equations of
motion are valid in any frame. The compensating terms provide
inertial forces depending on the total momentum ${\bf P}$, intrinsic
angular momentum ${\bf J}$ and intrinsic inertia tensor ${\bf I}$.
Therefore, the privileged frames where Newton's equations are valid
(\textit{Newtonian} frames) are completely determined by the matter
distribution of the universe (\textit{Machianization}). At the
Hamiltonian level, the gauge invariance leads to first class
constraints that remove those degrees of freedom that make no sense
once the absolute space has been eliminated. This reformulation of
classical mechanics is entirely relational, since it is a dynamics
for the distances between particles. It is also Machian, since the
rotation of the rest of the universe produces centrifugal effects.
It then provides a new perspective to consider the foundational
ideas of general relativity, like Mach's principle and the weak
equivalence principle. With regard to the concept of time, the
absence of an \textit{absolute time} is known to be a characteristic
of \textit{parametrized} systems. Furthermore, the scale invariance
of those parametrized systems whose potentials are inversely
proportional to the squared distances can be also gauged by
introducing another compensating term associated with the intrinsic
virial $G$ (\textit{shape-dynamics}).\vskip1.5cm
\end{abstract}

\maketitle

\section{Introduction}

Newton's mechanics describes particle motions relative to external
absolute space and time. The evolution of an isolated system of $N$
particles is determined by a set of independent initial positions
and velocities, whose sole ambiguity lies in the choice of the
inertial frame we use to express the initial configuration. Thus the
system has $3N$ degrees of freedom. The quality of \textit{inertial}
is conferred by the state of motion of the frame relative to the
absolute space. This approach was criticized by Leibniz and Mach,
who claimed for a dynamics of relations among particles without any
reference to external non-material entities \cite{Alexander,Mach}.
Mach's ideas guided Einstein in its way towards general relativity:
\textit{``... it is contrary to the mode of thinking in science to
conceive of a thing (the space-time continuum) which acts itself,
but which cannot be acted upon. This is the reason why E. Mach was
led to make the attempt to eliminate space as an active cause in the
system of mechanics''} \cite{Einstein}. During the last century,
different strategies have been tried to formulate a relational
mechanics in agreement with Mach's thought. Most of them were based
on a reformulation of the kinetic energy as a sum of ``interaction''
terms decreasing with the distance
\cite{Reissner,Schrodinger,Barbour75,Barbour77,Barbour95}. However
these approaches predicted non-observed anisotropies of the inertia
\cite{Barbour77}, and could not explain the undisputable success of
Newton's theory. The idea that the Lagrangian should exhibit gauge
invariance under the Galilean group was introduced in References
\onlinecite{Barbour77} and \onlinecite{Barbour82}. This requirement
guarantees the abolition of the absolute space, since any frame
becomes valid to apply the dynamical equations; but it does not
completely prescribe the Lagrangian. In Ref.~\onlinecite{Barbour82}
a given Lagrangian is \textit{gauged} by replacing its measure
(kinetic energy) with a measure defined in the space of
\textit{orbits} (each orbit represents a set of configurations that
are equivalent under gauge transformations). The measure between two
orbits, or \textit{intrinsic differential}, results from minimizing
the original measure between configurations belonging to the
involved orbits (``best matching'' \cite{Barbour03}). The so
obtained gauge invariant dynamics is left with those solutions to
the original Lagrangian having vanishing momentum and angular
momentum. This strategy has been improved in Ref.~\onlinecite{Gryb},
where it is extended to arbitrary symmetry groups and finite
transformations, on the basis of comparing with Yang-Mills gauge
theories. In this article we will follow the approach of gauging a
well established Lagrangian: the one for standard Newton's
mechanics. Instead of defining a measure between orbits, we will
introduce counterterms in the kinetic energy to compensate the
changes of Newton's kinetic energy under gauge transformations. In
this way, we will get an explicitly gauge invariant Lagrangian
leading to equations of motion that are valid in any frame.

\vskip1cm

Newton's laws are invariant under \textit{rigid} (time-independent)
time translations, and space translations and rotations of
particle's positions $\mathbf{r}_{i}$:%
\begin{eqnarray}
\text{time translation}\text{: \ \ \ \ \ \ \ \ \ \ \ \ \ \ \ }
&&t~\longrightarrow ~t~+~\epsilon ~,  \label{time} \\
\text{space translation: \, \ \ \ \ \ \ \ \ \ \ \ \ } &&\mathbf{r}%
_{i}~\longrightarrow ~\mathbf{r}_{i}~+~\mathbf{\xi }~,  \label{trans} \\
\text{rotation: \; \ \ \ \ \ \ \ \ \ \ \ \ \ \ \ \ \ \ \ \ \ \ \ } &&%
\mathbf{r}_{i}~\longrightarrow ~\mathbf{A\cdot r}_{i}~,  \label{rotation}
\end{eqnarray}%
where $\mathbf{A}$ is an orthogonal matrix. If the rotation is
infinitesimal, it becomes%
\begin{equation}
\mathbf{r}_{i}~\longrightarrow ~\mathbf{r}_{i}~+\mathbf{\alpha }~\times ~%
\mathbf{r}_{i}~,  \label{infinitrotation}
\end{equation}%
where $\mathbf{\alpha }$ is a vector directed along the axis of rotation,
whose infinitesimal modulus is the angle of rotation. Besides, Newton's laws
are invariant under Galileo transformations,%
\begin{equation}
\text{Galileo transformation: \ \ \ \ \ \ \ \ \ }\mathbf{r}%
_{i}~\longrightarrow ~\mathbf{r}_{i}~+~\mathbf{V}t~,  \label{boost}
\end{equation}%
which constitute a particular case of \textit{local }(time-dependent) space
translations: those having $\mathbf{\xi }(t)=\mathbf{V}t$ .

\bigskip

Transformations (\ref{time}-\ref{rotation}) and (\ref{boost}) constitute the
\textit{Galilean group} of Newton's mechanics. In Newton's mechanics,
absolute space and time are represented by the privileged family of inertial
frames and clocks, which relate each other through the transformations of
the Galilean group. The abolition of absolute space then calls for the
elimination of privileged frames: the \textit{dynamics of relations} must be
governed by laws to be applied in any frame. This goal can be attained by
building the Mechanics as a gauge theory of the Galilean group. To turn the
Galilean group into a gauge group means to succeed in becoming local those
already existing rigid symmetries. If so, not only rigid rotations will be
allowed but any time-dependent rotation too; not only Galileo
transformations will leave invariant the equations of motion but any
time-dependent space translation too. The absolute time would be eliminated
as well, since gauging the time translation implies that the parameter $t$
can be changed in an arbitrary way: $t~\longrightarrow ~t~+~\epsilon (t)$;
thus, the time parameter would become physically irrelevant. In sum, the
dynamics of relations calls for Lagrangians that are invariant under
arbitrary \textit{time-dependent} rotations and translations together with
arbitrary redefinitions of the time parameter. In such case, no privileged
frames and clocks will remain in the formulation of mechanics, what is the
mark of Relational Mechanics. Since Galileo transformations will be subsumed
in the subgroup of local translations, it results that the gauge group of
Relational Mechanics is composed of transformations (\ref{time}-\ref%
{rotation}).\ This gauge group has been called \textit{Leibniz group} in
Ref.~\onlinecite{Barbour77} (see also Ref.~\onlinecite{Ehlers}).

\bigskip

A typical feature of gauge theories is the existence of constraints
among the canonical variables. This means that not all the variables
represent genuine degrees of freedom. In particular, by gauging
translations and rotations the system of $N$ particles is left with
$3N-6$ degrees of freedom. The dynamics of relations has less
degrees of freedom than Newton's dynamics because evolutions
consisting in rigid translations and rotations means nothing in
Relational Mechanics. The redundant variables can be frozen by
fixing the gauge. In Relational Mechanics this procedure is
equivalent to choose a frame. As we will show, Relational Mechanics
contains frames where Newton's equations are valid. But such frames
are determined by the matter distribution of the universe, as
required by Mach.

\bigskip

In Section \ref{gauging} we obtain a Lagrangian that is invariant
under local translations and rotations. We study the constraints and
gauge fixing conditions associated with this Lagrangian. In Section
\ref{equations} we obtain the relational equations of motion, and
show the existence of frames where Newton's equations are valid. We
apply these equations to the case of Newton's bucket to show that
the centrifugal effect would disappear if the rest of the universe
were absent. We compare this result with the nature of the
centrifugal effect in general relativity. We also discuss the status
of the weak equivalence principle. In Section \ref{elimination} we
explain how the absolute time is eliminated by parametrizing the
system. In Section \ref{scaleinvariance} we show that the scale
invariance can also be gauged by including a counterterm associated
with the intrinsic virial function (the scale invariance requires
potentials that are inversely proportional to the squared
distances). In Section \ref{conclusions} we display the conclusions.

\section{Gauging translations and rotations}

\label{gauging}

As a first step in the way to its relational formulation, Mechanics should
be reduced to a dynamics of relative positions%
\begin{equation}
\mathbf{r}_{ij}~\doteq ~\mathbf{r}_{i}-\mathbf{r}_{j}~.
\end{equation}%
This goal can be reached by \textit{gauging} the translations; this means
the process of becoming \textit{local} (dependent on time) the rigid
symmetry (\ref{trans}). In fact, velocities are not invariant under local
translations but change as%
\begin{equation}
\mathbf{v}_{k}~\longrightarrow ~\mathbf{v}_{k}+{\dot{\mathbf{\xi }}}(t)~.
\end{equation}%
Therefore, a Lagrangian invariant under local translations is expected to
depend just on relative velocities $\mathbf{v}_{ij}={\dot{\mathbf{r}}}_{ij}$.

\bigskip

The process of gauging a rigid symmetry in gauge theories involves
the introduction of gauge fields to compensate the (bad) behavior of
derivatives under local transformations. In a more technical
language, one introduces a \textit{connection} to make the
derivatives behave in a \textit{covariant} way; i.e., thanks to the
compensating terms the derivatives behave in the same way both under
local and rigid transformations. In gauge theory, the dynamics is
constrained by relations emerging from the local symmetry the fields
obey. The constrains imply that not all the fields are genuine
degrees of freedom, since their initial values cannot be chosen in a
completely free way. A remarkable feature of gauge fields is that
they mediate interactions by carrying energy-momentum at a finite
propagation velocity. Instead, classical mechanics describes
interactions through potentials $V(r_{ij})$ depending just on the
distances between particles; it is the realm of {\it instantaneous
interactions at a distance}. As a consequence, the degrees of
freedom associated with the external space and time can be
eliminated \textit{without} introducing new fields, but just
resorting to the same dynamical variables we started from. This can
be made by providing the Lagrangian with compensating terms
depending on positions and velocities. Notice that $V(r_{ij})$ is
already a gauge invariant quantity. Therefore, mechanics will become
{\it relational} if the kinetic energy in the Lagrangian becomes a
gauge invariant quantity. Newton's kinetic energy is not invariant
under local translations $\mathbf{\xi }(t)$; if $\mathbf{\xi }$ is
infinitesimal, it changes as
\begin{equation}
\sum\limits_{k}\frac{m_{k}}{2}~\delta (\mathbf{v}_{k}\cdot \mathbf{v}%
_{k})~=\sum\limits_{k}m_{k}~\mathbf{v}_{k}\cdot \delta \mathbf{v}_{k}~=~{%
\dot{\mathbf{\xi }}}\cdot \sum\limits_{k}m_{k}~\mathbf{v}_{k}=~{\dot{\mathbf{%
\xi }}}\cdot \mathbf{P}~.
\end{equation}%
Remarkably, the square total momentum $\mathbf{P}$ changes essentially in
the same way:%
\begin{equation}
\delta (\mathbf{P}\cdot \mathbf{P})~=~2~\mathbf{P}\cdot \delta \mathbf{P}=~2~%
\mathbf{P}\cdot ~\sum\limits_{k}m_{k}~{\dot{\mathbf{\xi
}}}~=~2M{\dot{\mathbf{\xi }}}\cdot \mathbf{P}~.
\end{equation}%
So, we can gauge the translations by compensating the kinetic energy with
the counterterm $\mathbf{P}\cdot \mathbf{P}/(2M)$. The \textit{gauged}
kinetic energy turns out to be the kinetic energy in a frame where the
center of mass is at rest; it gets the form of an \textit{intrinsic} kinetic
energy: \footnote{%
Intrinsic quantities have the form $\sum\limits_{i<j}\frac{m_{i}m_{j}}{2M}%
~f_{ij}(\mathbf{r}_{ij},\mathbf{v}_{ij})$ where $f_{ij}=f_{ji\,}$.}
\begin{equation}
\sum\limits_{i<j}\frac{m_{i}m_{j}}{2M}~\mathbf{v}_{ij}\cdot \mathbf{v}%
_{ij}~=\sum\limits_{k}\frac{m_{k}}{2}~\mathbf{v}_{k}\cdot \mathbf{v}_{k}~-~%
\frac{\mathbf{P}\cdot \mathbf{P}}{2M}~=\sum\limits_{k}\frac{m_{k}}{2}%
~\left( \mathbf{v}_{k}-\frac{\mathbf{P}}{M}\right) \cdot \left( \mathbf{v}%
_{k}-\frac{\mathbf{P}}{M}\right) ~.  \label{kinetic}
\end{equation}%
So, the Lagrangian%
\begin{equation}
L~=~\sum\limits_{i<j}\,\frac{m_{i}~m_{j}}{2M}~\mathbf{v}_{ij}\cdot \mathbf{v}%
_{ij}~-~V(r_{ij})~  \label{lagrangian}
\end{equation}%
is invariant under local translations. Remarkably, in the
kinetic energy (\ref{kinetic}) the derivative of the position $\mathbf{v}%
_{k}=d\mathbf{r}_{k}/dt$ is accompanied by a compensating term $\mathbf{P}/M$%
, in such a way that the complete expression is invariant not only under
rigid translations but under local translations too. In fact,%
\begin{equation}
\mathbf{v}_{k}-\frac{\mathbf{P}}{M}\longrightarrow \mathbf{v}_{k}+{\dot{%
\mathbf{\xi }}}-\frac{\mathbf{1}}{M}\sum\limits_{i}m_{i}~(\mathbf{v}_{i}+{%
\dot{\mathbf{\xi }})}=\mathbf{v}_{k}-\frac{\mathbf{P}}{M}~.
\end{equation}%
It could be said that the compensating term $\mathbf{P}/M$ \textquotedblleft
covariantizes\textquotedblright\ the time derivative under local
translations.

\bigskip

We will follow the same strategy for gauging the rotations. Lagrangian (\ref%
{lagrangian}) is invariant under rigid rotations. Instead, local
infinitesimal rotations produce the change
\begin{equation}
\delta (\mathbf{v}_{ij}\cdot \mathbf{v}_{ij})~=~2~\mathbf{v}_{ij}\cdot
\delta \mathbf{v}_{ij}=~2~\mathbf{v}_{ij}\cdot \frac{d}{dt}(\mathbf{\alpha }%
~\times ~\mathbf{r}_{ij})=~2~\mathbf{v}_{ij}\cdot (\dot{\mathbf{\alpha }}%
\times ~\mathbf{r}_{ij})~=~2~\dot{\mathbf{\alpha }}\cdot (\mathbf{r}%
_{ij}~\times ~\mathbf{v}_{ij})~,
\end{equation}%
which makes the kinetic energy vary in
\begin{equation}
\sum\limits_{i<j}\,\frac{m_{i}~m_{j}}{2M}~\delta (\mathbf{v}_{ij}\cdot
\mathbf{v}_{ij})~=~\dot{\mathbf{\alpha }}\cdot ~\mathbf{J}~,  \label{varyT}
\end{equation}%
where $\mathbf{J}$ \ is the \textit{intrinsic} angular momentum or spin,
which is invariant under local translations:
\begin{equation}
\mathbf{J}~\ \doteq ~\ \sum\limits_{i<j}\,\frac{m_{i}~m_{j}}{M}~\mathbf{r}%
_{ij}\times \mathbf{v}_{ij}~=~\sum\limits_{k}\,m_{k}~\mathbf{r}_{k}\times
\mathbf{v}_{k}-\mathbf{R}\times \mathbf{P}~=~\sum\limits_{k}\,m_{k}~(\mathbf{%
r}_{k}-\mathbf{R})\times \left( \mathbf{v}_{k}-\frac{\mathbf{P}}{M}\right) ~
\label{J}
\end{equation}%
($\mathbf{R}$ is the center-of-mass position). To compensate the behavior (%
\ref{varyT}) we will provide the kinetic energy with a counterterm quadratic
in $\mathbf{J}$. This term has to be invariant under rigid translations and
rotations, so we write it as $(1/2)\,\mathbf{J}\cdot \mathbf{I}^{-1}\cdot
\mathbf{J}$, where $\mathbf{I}^{-1}$ is some symmetric tensor depending only
on relative positions (then, $\dot{\mathbf{\alpha }}$ does not appear in the
local rotation of $\mathbf{I}^{-1}$). Under local rotations the counterterm
varies in
\begin{equation}
\frac{1}{2}~\delta (\mathbf{J}\cdot \mathbf{I}^{-1}\cdot \mathbf{J)~}\ =\
~\delta \mathbf{J}\cdot \mathbf{I}^{-1}\cdot \mathbf{J}~,
\label{varycounterrotation}
\end{equation}%
where we only use those terms in $\delta \mathbf{J}$ that are
proportional to $\dot{\mathbf{\alpha }}$, since rigid rotations do
not change the counterterm:
\begin{equation}
\delta \mathbf{J}\ =\ \sum\limits_{i<j}\frac{m_{i}~m_{j}}{M}~\mathbf{r}%
_{ij}~\times (\dot{\mathbf{\alpha }}\times ~\mathbf{r}_{ij})+...
\end{equation}%
Therefore
\begin{equation}
\delta \mathbf{J}\ =\ \sum\limits_{i<j}\frac{m_{i}~m_{j}}{M}~[r_{ij}^{2}~%
\dot{\mathbf{\alpha }}-\mathbf{r}_{ij}~(\dot{\mathbf{\alpha }}\cdot \mathbf{r%
}_{ij})]~\ +\ ...~\ =\ ~\dot{\mathbf{\alpha }}\cdot \mathbf{I}\ +\ ...
\label{varyJ}
\end{equation}%
where $\mathbf{I}$ is the \textit{intrinsic }inertia tensor,%
\begin{eqnarray}
\mathbf{I}~ &\doteq&~\sum\limits_{i<j}\frac{m_{i}~m_{j}}{M}~[r_{ij}^{2}~\mathbf{1}%
-\mathbf{r}_{ij}\otimes \mathbf{r}_{ij}]\ =~\sum\limits_{k}m_{k}\,(r_{k}^{2}~%
\mathbf{1}-\mathbf{r}_{k}\otimes \mathbf{r}_{k})-M\,(R^{2}~\mathbf{1}-%
\mathbf{R}\otimes \mathbf{R})\   \notag \\
\ &=&~\sum\limits_{k}m_{k}\,\left[ |\mathbf{r}_{k}-\mathbf{R}|^{2}~\mathbf{1}%
-(\mathbf{r}_{k}-\mathbf{R})\otimes (\mathbf{r}_{k}-\mathbf{R})\right] ~,
\label{Itensor}
\end{eqnarray}%
which is just the inertia tensor relative to the center of mass. Replacing
the result (\ref{varyJ}) in Eq.~(\ref{varycounterrotation}) we reproduce the
behavior (\ref{varyT}) of the intrinsic kinetic energy under local
rotations. Therefore, not only translations but also rotations are gauged if
the Lagrangian is taken to be
\begin{equation}
L~=~\sum\limits_{i<j}\frac{m_{i}~m_{j}}{2M}~\mathbf{v}_{ij}\cdot \mathbf{v}%
_{ij}~-~\frac{1}{2}~\mathbf{J}\cdot \mathbf{I}^{-1}\cdot \mathbf{J}%
~-~V(r_{ij})~.  \label{newlagrangian}
\end{equation}%
Lagrangian (\ref{newlagrangian}) was introduced by Lynden-Bell in
References \onlinecite{Lynden92,Lynden95,Katz95}; the counterterms
were there obtained by looking for the minimum value the kinetic
energy can reach under finite local translations and rotations. Such
a minimum is reached for the
values $\dot{\mathbf{\xi }}=\mathbf{P}/M$ and $\dot{\mathbf{\alpha }}=%
\mathbf{I}^{-1}\cdot \mathbf{J}$. In fact, it is remarkable that Lagrangian (%
\ref{newlagrangian}) can be rewritten as%
\begin{equation}
L~=~\sum\limits_{k}\frac{m_{k}}{2}~\left\vert \mathbf{v}_{k}-\frac{\mathbf{P}%
}{M}-(\mathbf{I}^{-1}\cdot \mathbf{J})\times (\mathbf{r}_{k}-\mathbf{R}%
)\right\vert ^{2}-~V(r_{ij})~  \label{newlagrangian2}
\end{equation}%
(Hints: use $\sum\limits_{k}m_{k}~\mathbf{r}_{k}\times \mathbf{v}_{k}-%
\mathbf{R}\times \mathbf{P}=\mathbf{J}$; prove $\sum\limits_{i<j}m_{i}m_{j}~%
\mathbf{r}_{ij}\otimes \mathbf{r}_{ij}~/M=\sum\limits_{k}m_{k}~(\mathbf{r}%
_{k}-\mathbf{R})\otimes (\mathbf{r}_{k}-\mathbf{R})$). Although
Lagrangian (\ref{newlagrangian}) describes the dynamics in terms of
relative positions and velocities, it is still a function of
individual positions and velocities because the relative positions
cannot be considered as independent variables.

\subsection{Constraints}

Gauge invariance implies constraints among the momenta. The conjugate
momenta coming from the Lagrangian (\ref{newlagrangian}) are%
\begin{equation}
\mathbf{p}_{k}\ =\ m_{k}\ \left[ \mathbf{v}_{k}-\frac{\mathbf{P}}{M}-(%
\mathbf{I}^{-1}\cdot \mathbf{J})\times (\mathbf{r}_{k}-\mathbf{R})\right]
\label{newmomentum}
\end{equation}%
(Hint: in Eq.~(\ref{newlagrangian}) use the result $\partial J_{\mu
}/\partial v_{k}^{\nu }=\epsilon _{\mu \lambda \nu }~m_{k}\ (\mathbf{r}_{k}-%
\mathbf{R})^{\lambda }$). Then, the momenta look like conventional
momenta for particles whose velocities have been diminished by the
motion of a rigid body that accompanies the center of mass and
rotates with a time dependent angular velocity $\mathbf{I}^{-1}\cdot
\mathbf{J}$, which is the mean angular velocity of the system.
Moreover, $\mathbf{p}_{k}$ behaves as a vector not only under rigid
rotations but under local rotations too; such a property is
guaranteed by the compensating terms in Eq.~(\ref{newmomentum}) (to
prove it, use Eq.~(\ref{varyJ})). \footnote{We can then say that
$\mathbf{p}_k = m_k\, D(\mathbf{r}_k-\mathbf{R})/Dt$, where the
\textit{covariant} derivative is defined as
$D\mathbf{r}_{ij}/Dt\doteq d\mathbf{r}_{ij}/dt-(\mathbf{I}^{-1}\cdot
\mathbf{J})\times \mathbf{r}_{ij~}$.}

\bigskip

The momenta accomplish the constraint equations
\begin{equation}
\mathcal{P}\,\doteq \,\sum\limits_{k}~\mathbf{p}_{k}\,=\,0~,\hspace{0.7in}%
\mathcal{J}\,\doteq \,\sum\limits_{k}~\mathbf{r}_{k}\times \mathbf{p}%
_{k}\,=\,0~.  \label{constraints}
\end{equation}%
The first one is trivial because $\sum m_{k}(\mathbf{v}_{k}-\mathbf{P}/M)=0$
and $\sum m_{k}(\mathbf{r}_{k}-\mathbf{R})=0$. To prove the second one, we
will use that $m_{k}(\mathbf{r}_{k}-\mathbf{R})=\sum\limits_{j}m_{k}m_{j}~%
\mathbf{r}_{kj}/M$; then%
\begin{eqnarray}
\mathcal{J\ } &=&\mathcal{\ }\sum\limits_{k}~\mathbf{r}_{k}\times \mathbf{p}%
_{k}\mathcal{\ }=\mathcal{\ }\sum\limits_{k}m_{k}~\mathbf{r}_{k}\times \left[
\mathbf{v}_{k}-\frac{\mathbf{P}}{M}-(\mathbf{I}^{-1}\cdot \mathbf{J})\times (%
\mathbf{r}_{k}-\mathbf{R})\right]  \notag \\
&&  \notag \\
&=&\mathcal{\ }\sum\limits_{k}m_{k}~\mathbf{r}_{k}\times \mathbf{v}_{k}\ -\
\mathbf{R}\times \mathbf{P}\ -\ \sum\limits_{k,\ j}\frac{m_{k}\,m_{j}}{M}~%
\mathbf{r}_{k}\times \left[ (\mathbf{I}^{-1}\cdot \mathbf{J})\times \mathbf{r%
}_{kj}\right] \mathcal{\ }=\mathcal{\ }\mathbf{J}\ -\ \sum\limits_{i<j}\frac{%
m_{i}\,m_{j}}{M}~\mathbf{r}_{ij}\times \left[ (\mathbf{I}^{-1}\cdot \mathbf{J%
})\times \mathbf{r}_{ij}\right]  \notag \\
&&  \notag \\
&=&\ \mathbf{J}\ -\ \sum\limits_{i<j}\frac{m_{i}~m_{j}}{M}~\left[ (\mathbf{I}%
^{-1}\cdot \mathbf{J})~\mathbf{r}_{ij}\cdot \mathbf{r}_{ij}-\mathbf{r}%
_{ij}~~[\mathbf{r}_{ij}\cdot (\mathbf{I}^{-1}\cdot
\mathbf{J})]\right] \ =\ \mathbf{J}\ -\ \mathbf{I}\cdot
(\mathbf{I}^{-1}\cdot \mathbf{J})\ =\ 0
\end{eqnarray}

\subsection{Gauge fixing conditions}

\label{gaugefixing}

Noether's theorem guarantees that vectors $\mathcal{P}$ and
$\mathcal{J}$ are conserved, because of the (rigid) symmetry under
translations and rotations. This means that the constraints are
compatible with the evolution. On the other hand,
the constrained quantities $\mathcal{P}$ and $\mathcal{J}$ are the \textit{%
generators} of translations and rotations, i.e. those symmetries we are just
gauging. In fact, the effects of translations and infinitesimal rotations on
the canonical variables are%
\begin{eqnarray}
\text{space translation:}\ \ \ \ \ \ \ \ \ \ \ \ \ \ \ \ \ \ \ \delta
x_{i}^{\mu } &=&\xi ^{\nu }\mathbf{\,\{}x_{i}^{\mu },\mathcal{P}_{\nu }\}~,%
\hspace{0.3in}\delta p_{i}^{\mu }\ =\ ~\xi ^{\nu }\mathbf{\,\{}p_{i}^{\mu },%
\mathcal{P}_{\nu }\},  \label{canonicalp} \\
&&  \notag \\
\text{infinitesimal rotation: \thinspace\ \ \ \ \ \ \ \ \ \ \ \ }\delta
x_{i}^{\mu } &=&\alpha ^{\nu }\mathbf{\,\{}x_{i}^{\mu },\mathcal{J}_{\nu }\},%
\hspace{0.3in}\delta p_{i}^{\mu }\ =\ \alpha ^{\nu }\mathbf{\,\{}p_{i}^{\mu
},\mathcal{J}_{\nu }\},  \label{canonicalJ}
\end{eqnarray}%
where Greek indices stand for Cartesian components (sums over
repeated Greek indices are assumed). $\mathcal{P}_{\mu \,},
\mathcal{J}_{\nu }$ satisfy the algebra
\begin{equation}
\{\mathcal{P}_{\mu \,},\mathcal{P}_{\nu }\}=0\,,\hspace{0.4in}\{\mathcal{P}%
_{\mu \,},\mathcal{J}_{\nu }\}=\epsilon _{\mu \nu \lambda }\ \mathcal{P}%
^{\lambda }\,,\hspace{0.4in}\{\mathcal{J}_{\mu \,},\mathcal{J}_{\nu
}\}=\epsilon _{\mu \nu \lambda }\ \mathcal{J}^{\lambda }\,.
\label{algebra}
\end{equation}

In the language of Dirac's formalism for constrained Hamiltonian systems
\cite{Dirac}, Eq.~(\ref{algebra}) says that the constraints $\mathcal{%
P_{\mu }},~\mathcal{J_{\nu }}$ are first class: they commute on the
constraint surface, besides of being compatible with the evolution.
In Dirac's formalism, first class constraints are the generators of
gauge transformations. Each gauge freedom implies a spurious degree
of freedom (a degree of freedom that is not determined by the
dynamics). The spurious degrees of freedom can be frozen by fixing
the gauge. In our case, the six first class constraints
$\mathcal{P_{\mu }},~\mathcal{J_{\nu }}$ imply that a system of $N$
particles has $3N-6$ genuine degrees of freedom. In the simpler
cases, one fixes the gauge by freezing quantities that are canonical
conjugated to the constraints. To show it, let us consider the Lagrangian $L(%
\dot{q},\dot{Q})=\dot{q}^{2}/2+\dot{Q}$, that leads to the constraint $P=1$.
This constraint is compatible with the evolution, since Lagrange equations
say that $dP/dt=0$. However this equation does not explain how the variable $%
Q$ evolves, because $P$ is not a function of $\dot{Q}$. Since
Lagrange equations do not determine the dynamics of $Q$, then $Q$
can be frozen through a gauge condition. At the Hamiltonian level,
it is not possible to write \thinspace $H=\dot{Q}P+\dot{q}p-L$ as a
function of the canonical variables $q,\,Q,\,p,\,P$, because the
formalism is unable to provide the function $\dot{Q}(q,Q,p,P)$. So,
the first term of $H$ is the constraint times an undetermined
function. Thus, this trivial example gives an idea of why
the Hamiltonian of a system harboring first class constraints $\mathcal{G}%
_{b}=0$ must contain terms proportional to $\mathcal{G}_{b}$ with
undetermined coefficients $\lambda ^{a}(t)$:%
\begin{equation}
H^{\prime }=H+\sum_{a}\ \lambda ^{a}(t)\,\mathcal{G}_{a}~.  \label{Hprime}
\end{equation}%
The ambiguity associated with the functions $\lambda ^{a}(t)$ does
not affect the evolution of gauge invariant quantities ({\it
observables}), since they commute with the $\mathcal{G}_{a}$'s on
the constraint surface (they are not affected by gauge
transformations). On the other side, the terms $\lambda
^{a}(t)\,\mathcal{G}_{a}$ imply the existence of quantities whose
evolution is not determined by the system: those quantities that do
not commute with the $\mathcal{G}_{a}$'s on the constraint surface.
Such gauge freedom can be
frozen by fixing the gauge. In general, a set of gauge fixing conditions $%
C_{a}=0$ will be admissible for a set of first class constraints $\mathcal{G}%
_{b}=0$ if they satisfy the requirement $\det \{C_{a},\mathcal{G}_{b}\}\neq
0 $ \cite{Henneaux}.

\bigskip

In Relational Mechanics fixing the gauge\textit{\ means choosing the
frame}. We can choose a frame by fixing the origin of coordinates at
the center of mass (but any other way of fixing the motion of the
center of mass is feasible), and fixing the orientation through some
convenient criterion. For instance, we could fix the three Cartesian
products of inertia $I_{\mu \nu }$, $\mu \neq \nu$, to be zero (or
any other way of fixing the products of inertia as
functions of time). In such case, the set of gauge conditions is $%
C_{a}=X_{\mu \,},\ I_{\mu \nu }$, where $X_{\mu }$ are the Cartesian
components of $\mathbf{R}$. The matrix $\{C_{a},\mathcal{G}_{b}\}$ has the
elements
\begin{equation}
\{X_{\lambda },\,\mathcal{P}_{\mu }\}\ =\ \delta _{\lambda \mu }\,,\hspace{%
1in}\{X_{\lambda },\,\mathcal{J}_{\mu }\}\ =\ \epsilon _{\lambda \mu
\nu }\,\,X^{\nu }\,,\hspace{1in}\{I_{\lambda \mu
},\,\mathcal{P}_{\nu }\}\ =\ 0~,
\end{equation}%
together with the brackets $\{I_{\lambda \mu },\,\mathcal{J}_{\nu }\}$
arranged in the following table:%
\begin{equation}
\begin{tabular}{c|ccc}
& $\ \ \mathcal{J}_{x}\ \ $ & $\ \ \mathcal{J}_{y}\ \ $ & $\ \ \mathcal{J}%
_{z}\ \ $ \\ \hline
$I_{yz}$ & $I_{yy}-I_{zz}$ & $-I_{xy}$ & $I_{zx}$ \\
$I_{zx}$ & $I_{xy}$ & $I_{zz}-I_{xx}$ & $-I_{yz}$ \\
$I_{xy}$ & $-I_{zx}$ & $I_{yz}$ & $I_{xx}-I_{yy}$%
\end{tabular}%
\ \ \ +[\text{terms vanishing for }\mathbf{R}=0].
\end{equation}%
There is a problem in the case of degenerate principal axes of
inertia, since it would result $\det \{C_{a},\mathcal{G}_{b}\}=0$.
Let us consider this issue for a two-particles system. If the
$x-$axis is chosen
along the direction joining the particles, then it is $I_{xx}=0$ and $%
I_{yy}=I_{zz}$. The symmetry around the $x-$axis is inherent in this
system; the rotation around the $x-$axis cannot be considered as a
possible motion of the system. Then, we should not attempt to gauge
the rotation around the $x-$axis; $\mathcal{J}_{x}$ should not be
considered as a generator of a gauge transformation. Therefore,
$\mathcal{J}_{x}$ and $I_{yz}$ must be eliminated in the former
algebra, what leads to a non-null value for $\det
\{C_{a},\mathcal{G}_{b}\}$. Thus we have $5$ first class
constraints; so a two-particles system has only $1$ degree of
freedom.

\section{Equations of motion}

\label{equations}

The equations of motion deriving from the Lagrangian
(\ref{newlagrangian}) (see Eq.~(\ref{newmomentum}) as well) are
\footnote{As results from the Lagrangian formalism, $\mathbf{J}$ has
to be regarded as a function of $\mathbf{r}_{k}$'s and
$\mathbf{v}_{k}$'s. $\mathbf{\nabla }_{k}$ differentiates with
respect to $\mathbf{r}_{k}$ at fixed velocities.}
\begin{equation}
m_{k}~\frac{d}{dt}\left[ \mathbf{v}_{k}-\frac{\mathbf{P}}{M}-(\mathbf{I}%
^{-1}\cdot \mathbf{J})\times (\mathbf{r}_{k}-\mathbf{R})\right] ~=~-\mathbf{%
\nabla }_{k}V-\mathbf{\nabla }_{k}\left( \frac{1}{2}\ \mathbf{J}\cdot
\mathbf{I}^{-1}\cdot \mathbf{J}\right) ~.  \label{motion}
\end{equation}%
These equations describe motions of particles relative to a rigid
body that accompanies the center of mass and rotates with the mean
angular velocity of the entire isolated system (the universe). Each
particle $m_{k}$ is acted by two types of forces. On the one hand
the forces coming from the interaction potentials $V_{ij}(r_{ij})$;
near masses dominate these gauge invariant forces. On the other
hand, the distant masses act mainly through a gauge dependent force
associated with the total angular momentum of the universe. The
equations of motion are necessarily covariant under local
translations and rotations of the frame, since they come from an
invariant Lagrangian. In fact, the invariance under local
translations is evident. Besides, due to
the beneficial effect of the compensating terms, no terms proportional to $%
\dot{\mathbf{\alpha }}$ are left inside the square brackets after a local
rotation ($\delta \mathbf{p}_{k}=\mathbf{\alpha }(t)\times \mathbf{p}_{k}$).
However, a factor $\dot{\mathbf{\alpha }}$ will still appear because of the
time derivative of $\mathbf{p}_{k}$. This unwanted contribution will be
compensated by the behavior of the gauge dependent force at the right hand
side. As in Newton's mechanics, the equations (\ref{motion}) have to be
endowed with a set of initial conditions to determine a specific evolution.
However, the equations (\ref{motion}) do not distinguish between
configurations connected by a gauge transformation. In the eyes of equations
(\ref{motion}) two set of initial data differing in a gauge transformation,%
\begin{equation}
\delta \mathbf{r}_{k}=\mathbf{\xi }+\mathbf{\alpha }\times \mathbf{r}_{k}~,%
\hspace{0.3in}\delta \mathbf{v}_{k}=\mathbf{\dot{\xi}}+\mathbf{\alpha }%
\times \mathbf{v}_{k}+\mathbf{\dot{\alpha}}\times \mathbf{r}_{k}~,
\end{equation}%
are \textit{equivalent}. The number of degrees of freedom decreases
to $3N-6$ because the initial conditions of the arbitrary vector
functions $\mathbf{\xi }(t)$, $\mathbf{\alpha }(t)$ must be
subtracted.

\bigskip

Remarkably, Eq.~(\ref{motion}) says that Newton's laws are valid in frames
where $\mathbf{P}$ is constant and the mean angular velocity $\mathbf{I}%
^{-1}\cdot \mathbf{J}$ vanishes. \footnote{$\mathbf{J}\cdot
\mathbf{I}^{-1}\cdot \mathbf{J}$ is a positive definite function.
Therefore its derivatives vanish at the minimum $\mathbf{J}=0$.} We
call these privileged frames \textit{Newtonian}. Newtonian frames
relate each other through rigid rotations and uniform translations.
Newton's mechanics was criticized by Mach because it contained
\textit{real} effects coming from the state of motion relative to
the absolute space. These effects canceled out in the inertial
frames; but non-inertial frames included effects such as centrifugal
and Coriolis forces whose magnitude depended on the acceleration and
rotation of the frame relative to the absolute space. Such effects
are recognizable in Eq.~(\ref{motion}) as well; however, they are
not determined by the relation between the frame and the absolute
space but by the distribution of matter in the chosen frame. In
fact, the gauge dependent force can be developed as%
\begin{equation}
-\mathbf{\nabla }_{k}\left( \frac{1}{2}\ \mathbf{J}\cdot \mathbf{I}%
^{-1}\cdot \mathbf{J}\right) ~=~m_{k}\ (\mathbf{I}^{-1}\cdot \mathbf{J}%
)\times \left( \mathbf{v}_{k}-\frac{\mathbf{P}}{M}\right) \ +\ \mathbf{f}%
_{k}\ ,
\end{equation}%
where%
\begin{equation}
f_{k\mu }~=~-\frac{1}{2}\ J^{\lambda }\,J^{\nu }\ \frac{%
\partial I_{\lambda \nu }^{-1}}{\partial x_{k}^\mu}~.
\end{equation}%
We will use that $I^{\alpha \gamma }\,I_{\gamma \beta }^{-1}=\delta
_{\beta }^\alpha $, so it is $I^{\alpha \gamma }\,\partial I_{\gamma
\beta }^{-1}=-I_{\gamma \beta
}^{-1}\,\partial I^{\alpha \gamma }$. Therefore, it results%
\begin{equation}
f_{k\mu }~=~-\frac{1}{2}\ I^{\lambda \gamma }\,I_{\gamma \beta }^{-1}\ J^{\beta }\,J^{\nu }\ \frac{%
\partial I_{\lambda \nu }^{-1}}{\partial x_{k}^\mu }~=~\frac{1}{2}%
\ I_{\lambda \nu }^{-1}\,J^{\nu
}\ I_{\gamma \beta }^{-1}\ J^{\beta }\ \frac{\partial I^{\lambda \gamma }}{%
\partial x_{k}^\mu }~.
\end{equation}%
Since%
\begin{equation}
\frac{\partial I^{\lambda \gamma }}{\partial x_{k}^\mu }~=~m_{k}\
\left[ 2\,(x_{k\mu }-X_{\mu })\ \delta ^{\lambda \gamma }
-(x_{k}^\gamma -X^{\gamma })\ \delta _{\mu }^\lambda -(x_{k
}^\lambda-X^{\lambda })\ \delta _{\mu}^\gamma \right] ~,
\end{equation}%
it can be concluded that $\mathbf{f}_{k}$ is a centrifugal force
\begin{equation}
\ \mathbf{f}_{k}~=~-m_{k}\ (\mathbf{I}^{-1}\cdot \mathbf{J})\times
\left[(\mathbf{I}^{-1}\cdot \mathbf{J})\times
(\mathbf{r}_{k}-\mathbf{R})\right] ~. \label{centrifugal}
\end{equation}%
The other inertial forces are also present. All of them are
determined by the distribution of mass of the isolated system. In
fact, by solving Eq.~(\ref{motion}) for the motion relative to the
center of mass, one gets
\begin{equation}
m_{k}~\frac{d}{dt}\left[ \mathbf{v}_{k}-\frac{\mathbf{P}}{M}\right] ~=~-%
\mathbf{\nabla }_{k}V\ +\ 2\,m_{k}\ (\mathbf{I}^{-1}\cdot \mathbf{J})\times
\left( \mathbf{v}_{k}-\frac{\mathbf{P}}{M}\right) \ +\ \mathbf{f}_{k}~+\ %
m_{k}~\left[ \frac{d}{dt}(\mathbf{I}^{-1}\cdot \mathbf{J})\right] \times (\mathbf{r%
}_{k}-\mathbf{R})~.  \label{relativetoCM}
\end{equation}
As said, the privileged Newtonian frames of Eq.~(\ref{motion}) are
selected by the physical system itself. This property realizes what
is called \textit{Machianization} in Ref.~\onlinecite{Friedman}.
Newton's laws are recovered in the frame where the universe has null
mean angular velocity and its center of mass moves uniformly
\cite{Lynden92,Lynden95}. They are only the result of a particular
way of fixing the gauge in the relational equations of motion. This
is the manner in which Relational Mechanics agree with Mach's
desideratum.

\bigskip

The existence of Newtonian frames does not mean that Relational
Mechanics is eventually equal to Newton's theory. This is because
the initial conditions in Newtonian frames are constrained to
satisfy the gauge choice $\mathbf{J}=0$; so we are left with a
family of solutions smaller than the respective one in Newton's
theory. For instance, when studying the motion of two isolated
particles one can fix the $x-$axis of a center-of-mass frame by
choosing the direction joining the particles. In this frame the
system has a vanishing angular momentum $\mathbf{J}$ (the frame is
Newtonian); thus, the equations (\ref{motion}) just describe a
two-body system with a radial motion governed by
\begin{equation}
\mu \ \ddot{r}~=~-\frac{dV}{dr}~,\label{radial}
\end{equation}%
where $\mu $ is the reduced mass and $r$ is the distance between the
particles. So, while Newton's theory allows a Keplerian two-particle
system to develop orbital motions where $r$ oscillates between
periastron and apastron, this possibility is ruled out in
Eq.~(\ref{radial}). The relational evolution of $r$ does not involve
a centrifugal potential because the orbital motion is meaningless in
a universe containing just two particles. We will come back on this
issue in the following subsection.

\begin{figure}[b]
\centering \includegraphics[width=7cm]{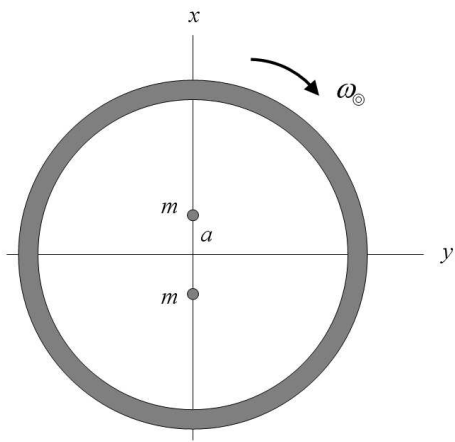} \caption{A
binary system surrounded by a shell representing the rest of the
universe.} \label{Fig1}
\end{figure}

\subsection{Newton's bucket and Mach's principle}

In Newton's mechanics the rotation relative to the absolute space
has physical consequences. Absolute rotation comes with a
centrifugal effect which could be verified by means of a turning
bucket filled with water. For Mach, instead, the parabolic shape of
the water surface is just the manifestation of the rotation relative
to the rest of the universe: \textit{``No one is competent to say
how the experiment would turn out if the sides of the vessel
increased in thickness and mass till they were ultimately several
leagues thick''} \cite{Mach}; i.e., nobody knows what would happen
if the mass of the bucket were comparable to the rest of the
universe or if the rest of the universe did not exist.

\bigskip

To study the issue in the light of equations (\ref{motion}), let us
consider a binary star composed of two equal stars at a distance
$2a$, orbiting its center of mass with circular motions of radius
$a$. The rest of the universe will be represented by a spherical
rigid shell centered at the center of mass of the system. The
dynamics can be alternatively analyzed in the center-of-mass frame
where the binary star is at rest but the shell rotates with angular
velocity $\omega _{\circledcirc }$ (see Figure \ref{Fig1}). In such
a frame the centrifugal force that equilibrates the binary star
comes from the rotation of the rest of the universe. Since the
chosen frame is made up of the principal axes of inertia of the
shell and the entire system, it results
\begin{equation}
\mathbf{J}=I_{\circledcirc }~\omega _{\circledcirc
}~\mathbf{e}_{z}~,\hspace{0.4in}\mathbf{J}\cdot
\mathbf{I}^{-1}=(\mathbf{I}^{-1})_{zz}~I_{\circledcirc }~\omega
_{\circledcirc }~\mathbf{e}_{z}~,
\end{equation}%
where $I_{\circledcirc }$ stands for the (isotropic) shell moment of
inertia. The chosen frame is clearly non-Newtonian, since
$\mathbf{J}\neq 0$. In Eq.~(\ref{relativetoCM}) let be $k=1$ the
label for one of the stars. Since $\mathbf{v}_{1}=0=\mathbf{P}/M$
and $\mathbf{I}^{-1}\cdot \mathbf{J}$ does not change in time, it
results
\begin{equation}
~\mathbf{f}_{1}~-\mathbf{\nabla }_{1}V~=~0~,~  \label{star}
\end{equation}%
where $-\mathbf{\nabla }_{1}V$ is the interaction force between the
stars (the spherical shell does not produces gravity in the inner
region). Equation~(\ref{star}) then says that the centrifugal force
due to the rotating shell equilibrates the gravitational attraction.
Since the inertia tensor $\mathbf{I}$ is diagonal in a frame of
principal axes of inertia, then it is
$(\mathbf{I}^{-1})_{zz}=1/I_{zz}$, where $I_{zz}=I_{\circledcirc
}+2\,m\,a^{2}$. \footnote{Eqs.~(\ref{J}) and (\ref{Itensor}) say
that $\mathbf{J}$ and $\mathbf{I}$ have the usual form in a frame
where $\mathbf{R}=0$. In particular they are additive.} Thus the
centrifugal force (\ref{centrifugal}) is
\begin{equation}
|\,\mathbf{f}\,|~=~\frac{m~\omega _{\circledcirc }^{2}~a}{\left(
1+\frac{2\,m\,a^{2}}{I_{\circledcirc }}\right)
^{2}}~.\label{centrif}
\end{equation}
Since $I_{\circledcirc }>>m~a^{2}$, Eq.~(\ref{star}) approaches the
form of the usual equation for orbital motion in Newton's mechanics.
However, if the rest of the universe were absent ($I_{\circledcirc
}=0$) then the centrifugal force would vanish and the proposed
orbital motion would make no sense, in agreement with Mach's ideas.

\bigskip

The geometry inside a rotating shell has been studied in general
relativity \cite{Brill,Pfister}. At the second order in the shell
angular velocity, the inner geometry is flat for a conveniently
prolate shell \cite{Pfister}. This means that any rotating chart
inside the shell will realize the usual inertial forces. However,
this general relativity result lacks the significance of
Eq.~(\ref{centrif}). In Eq.~(\ref{centrif}) the centrifugal force is
caused by the rotation of the shell relative to the stars; its
magnitude is proportional to the squared relative angular velocity.
If the rotation or the shell evaporated then the centrifugal force
would vanish and the binary star would collapse. In general
relativity, instead, the evaporation of the shell has no
consequences inside the shell: the geometry will continue to be
flat, and the binary star will continue orbiting. Actually the
general relativity solution possesses an inmaterial (``ethereal'')
second ``shell'' at the infinity, where the boundary conditions for
the metric are chosen.

\subsection{Weak equivalence principle}

The weak equivalence principle (WEP), one of the pillars of general
relativity, lies in the core of Newtonian mechanics. The Newtonian
motions of particles in an external gravitational potential do not
depend on their masses, what means that the inertial-gravitational
field can be locally canceled by changing to a freely falling frame.
This is the basis for the geometrization of freely falling motions.
Noticeably, we can hardly reproduce WEP in the context of Relational
Mechanics. In fact, the inertial forces in Eq.~(\ref{relativetoCM})
are not proportional to $m_k$, since $m_k$ is present in
$\mathbf{P}/M$ and $\mathbf{I}^{-1}\cdot \mathbf{J}$. Then, given a
self-gravitating system, we get that $m_k$ does take part in the
equation for $\mathbf{r}_k(t)$ (even so, its contribution can be
negligible for big systems, as seen in Eq.~(\ref{star}) where $m$
remains encapsuled in the ratio $2\,m\,a^2/I_{\circledcirc}$).
Instead, in Newtonian mechanics the inertial forces are proportional
to $m_k$ because the global properties of the isolated system, like
$\mathbf{I}^{-1}\cdot \mathbf{J}$, have been substituted for
absolute properties of the frame (absolute rotation, etc.).
Actually, WEP is a statement concerning individual motions of
particles. But the very idea of individual evolution means nothing
in Relational Mechanics, since ${\bf r}_k(t)$  can be completely
altered by a gauge transformation. In Relational Mechanics the
observables (gauge invariant magnitudes) are the distances between
particles. Instead, general relativity uses individual evolutions
(world-lines) and space-time geometry as the basic concepts.

\section{Elimination of the absolute time}

\label{elimination}

So far, translations and rotations have been gauged. We have obtained the
Lagrangian (\ref{newlagrangian}) which does not involve the idea of an
absolute space:\ it can be used in any frame. However it still depends on
the absolute time. To remove the absolute time we should gauge the time
translation $t\longrightarrow t+\epsilon $. Since $t$ is a parameter
identifying each configuration of the system in a temporal succession, a
time translation is a relabeling (\textit{reparametrization}) of the
consecutive configurations. The invariance of the Lagrangian under rigid
time translations is related to the conservation of the energy. Therefore,
we expect that the requirement of gauge (local) invariance under
reparametrizations will lead to a Hamiltonian first class constraint
\begin{equation}
\mathcal{H~}=~0~,
\end{equation}%
i.e. some relation among the momenta that causes the vanishing of
the Hamiltonian $H$.

\bigskip

Gauging the time translations does not involve counterterms for
compensating the bad behavior of derivatives in the kinetic energy,
since such behaviors do not come from the objects under
differentiation but from the operator $d/dt$ itself. In fact, under
local time translations $t\longrightarrow t+\epsilon
(t)$ the derivative changes as%
\begin{equation}
\frac{d}{dt}~\longrightarrow ~\frac{1}{1+\dot{\epsilon}(t)}\ \frac{d}{dt}~ \doteq ~%
\frac{1}{N(t)}\ \frac{d}{dt}~;  \label{derivative}
\end{equation}%
therefore velocities change as $\dot{q}\longrightarrow \lbrack 1+\dot{%
\epsilon}(t)]^{-1}\,\dot{q}=N(t)^{-1}\,\dot{q}$. Since the invariance of the
action is the invariance of $L\ dt$, then the behavior of the Lagrangian
under reparametrizations should be%
\begin{equation}
~L(q_{i},\,N^{-1}\ \dot{q}_{i})~=~N^{-1}\ L(q_{i},\,\dot{q}_{i})~.
\label{parameter}
\end{equation}%
This behavior characterizes the so called \textit{parametrized
systems }\cite{Lanczos,Sundermeyer}. The invariance of the action
means that the action is unable to distinguish the trajectories
$\mathbf{r}_{i}(t)$ from the reparametrized ones
$\mathbf{r}_{i}(t+\epsilon (t))$; therefore, the parameter $t$ is
physically irrelevant. Let us differentiate the equation
(\ref{parameter}) with respect to $N^{-1}$:
\begin{equation}
\sum\limits_{i}\ \dot{q}_{i}\ \frac{\partial L(q_{i},\,N^{-1}\ \dot{q}_{i})}{%
\partial (N^{-1}\ \dot{q}_{i})}~=~L(q_{i},\,\dot{q}_{i})~.
\end{equation}
Since this result is valid whatever the function $N(t)$ is, one can
substitute it with $N=1$ to conclude that parametrized systems come
with a constraint $\mathcal{H}$ that cancels out the Hamiltonian, as
expected. On the constraint surface, the Hamiltonian constraint
$\mathcal{H}$
commutes with $\mathcal{P}_{\mu \,}$, $\mathcal{J}_{\nu }\,$ because $%
\mathcal{P}_{\mu \,}$, $\mathcal{J}_{\nu }$ are conserved
quantities. Thus, the entire set of constraints remains first class.
In a parametrized system, the evolution along the irrelevant time
$t$ can be regarded as a gauge transformation, since the evolution
is carried on by the first class
Hamiltonian (see Eq.~(\ref{Hprime})) \footnote{%
We use the same notation $N(t)$ for the \textquotedblleft
free\textquotedblright\ functions in Eqs.~(\ref{derivative}) and (\ref%
{constraintH}) for reasons that will become clear in Eq.~(\ref{HE}).}
\begin{equation}
H^{\prime }=N(t)\ \mathcal{H}+\lambda ^{\mu }(t)\ \mathcal{P}%
_{\mu }+\chi ^{\nu }(t)\ \mathcal{J}_{\nu }~. \label{constraintH}
\end{equation}
As explained in Section \ref{gaugefixing}, the freedom associated
with $\mathcal{H}$ can be fixed by adding an admissible gauge
condition $C$. So we choose $C = \tau(q,p)-t$, where $\tau (q,p)$ is
invariant under translations and rotations, and satisfies
$\{\tau,\mathcal{H}\}>0$. Thus it results $\dot\tau=\{\tau,H'\}>0$,
which means that $\tau (q,p)$ is a dynamical quantity that
monotonically increases along the evolution; it is a
\textit{physical clock} defined by the system itself, also called
\textit{internal} time \cite{Kuchar,Beluardi}. The gauge condition
$C=0$ means that the irrelevant parameter $t$ can be replaced by the
internal time $\tau$.

\subsection{Hiding the time}

A well known example of parametrized system is the relativistic free
particle,
\begin{equation}
L\mathcal{~}=~-m\ \sqrt{\dot{Q}^{2}-\dot{q}^{2}}~, \label{rl}
\end{equation}%
which displays a constraint between the conjugate momenta:%
\begin{equation}
P=\frac{\partial L}{\partial \dot{Q}}=\frac{-m~\dot{Q}}{\sqrt{\dot{Q}^{2}-%
\dot{q}^{2}}}~,\hspace{0.3in}p=\frac{\partial L}{\partial \dot{q}}=\frac{m~%
\dot{q}}{\sqrt{\dot{Q}^{2}-\dot{q}^{2}}}~\hspace{0.3in}\Longrightarrow
\hspace{0.3in}P=\pm \sqrt{p^{2}+m^{2}}.
\end{equation}%
The constraint surface splits into two sheets; we will keep the negative
sign for $P$ ($\dot{Q}>0$). Then, the Hamiltonian constraint is%
\begin{equation}
\mathcal{H~}=\mathcal{~}P+\sqrt{p^{2}+m^{2}}~=~0~.
\end{equation}%
This constraint causes the vanishing of the Hamiltonian $H$:
\begin{equation}
H\mathcal{~}=\mathcal{~}P~\dot{Q}+p~\dot{q}-L\mathcal{~}=~\frac{\sqrt{\dot{Q}%
^{2}-\dot{q}^{2}}}{m}\ (-P^{2}+p^{2}+m^{2})~=~0~.\mathcal{~}
\end{equation}%
We can fix the gauge by choosing the physical (internal) time $\tau
=Q$ , that satisfies $\{\tau ,\mathcal{H}\}=1$. However this is not
the sole choice for the physical time. For instance, $\tau =Q\pm q$
yields $\{\tau,\mathcal{H}\}>0$; so, it is a good internal time. Any
function $\tau =f(Q)$ such that $f^{\prime }(Q)>0$ is a good choice
too. The gauge $Q-t=0$ fixes the irrelevant time parameter to be
equal to $Q$. So $\dot{Q}=1$ and the Lagrangian becomes $L =-m\,\int
\sqrt{1- {\dot q}^2}$. But, as said, there are many other forms for
replacing the parameter $t$ with an internal time.

\bigskip

The free relativistic particle also shows how to
\textit{parametrize} a system: we achieve the property
(\ref{parameter}) by replacing the velocities $\mathbf{v}_{k}$ with
$\mathbf{v}_{k}/\dot{Q}$ and multiplying the Lagrangian by a factor
$\dot{Q}$. Finally, we take $Q$ to be a new canonical variable. For
instance
\begin{equation}
L(\dot{q})\mathcal{~}=\mathcal{~}-m\sqrt{1-\dot{q}^{2}}~\longrightarrow ~L(%
\dot{q},\dot{Q})\mathcal{~}=\mathcal{~}-m\ \dot{Q}\ \sqrt{1-\frac{\dot{q}^{2}%
}{\dot{Q}^{2}}}~=~-m\ \sqrt{\dot{Q}^{2}-\dot{q}^{2}}~.
\end{equation}%
So, a possible strategy for eliminating the absolute time consists
in adding the system with a new canonical variable $Q$ causing the
parameter $t$ becomes physically irrelevant. The new variable does
not imply a new degree of freedom, since it comes together with a
new constraint (the Hamiltonian constraint); so the original number
of degrees of freedom of the non-parametrized system is kept. Once
the variable $Q$ was introduced through this procedure, it can be
mixed with the rest of canonical variables by means of canonical
transformations. It could be said that the procedure hides the
absolute time among the canonical variables. However, once the time
was hidden, there are many internal times --all of them on an equal
footing-- that can be retrieved by fixing the gauge. Thus, the
privileged absolute time was lost forever.

\bigskip

Let us parametrize a classical particle:
\begin{equation}
L(\mathbf{r},\mathbf{v},\dot{Q})~=\frac{1}{\dot{Q}}~\frac{m\ \mathbf{v}^{2}}{%
2\,}~-~\dot{Q}\ V~.  \label{classicalparametrized}
\end{equation}%
The canonical momenta are%
\begin{equation}
\mathbf{p}\ =\ \frac{m\,\mathbf{v}}{\dot{Q}}~,\hspace{0.3in}P\ =\ -\frac{m\
\mathbf{v}^{2}}{2\ \dot{Q}^{2}}-V=-\frac{\mathbf{p}^{2}}{2\,m}-V~.
\label{classicalP}
\end{equation}%
Then the Hamiltonian constraint is%
\begin{equation}
\mathcal{H~}=\mathcal{~}P+\frac{\,\mathbf{p}^{2}}{2\,m}+V~=~0~,
\label{classicalH}
\end{equation}%
and the Hamiltonian $H=P~\dot{Q}+\mathbf{p}\cdot \mathbf{v}-L$ vanishes. The
Hamiltonian evolution is governed by $H^{\prime }=N(t)\,\mathcal{H}$. Then
it is%
\begin{equation}
\mathbf{\dot{r}}\ =\ N(t)\,\{\mathbf{r},\mathcal{H}\}\ =\ N(t)\,\frac{%
\mathbf{p}}{m}~,\hspace{0.3in}\hspace{0.3in}\mathbf{\dot{p}}\ =\ N(t)\,\{%
\mathbf{p},\mathcal{H}\}\ =\ -N(t)\,\mathbf{\nabla }V~,  \label{HE}
\end{equation}%
whose solution is expressed in terms of an arbitrary time parameter $\int
N(t)\,dt$ that can be freely fixed. Besides it is%
\begin{equation}
\dot{Q}\ =\ N(t)\,\{Q,\mathcal{H}\}\ =\ N(t)~,\hspace{0.3in}\hspace{0.3in}%
\dot{P}\ =\ N(t)\,\{P,\mathcal{H}\}\ =\ 0~.\label{dotQ}
\end{equation}%
Equation (\ref{classicalP}) shows that the conserved quantity $P$ is
(minus) the energy of the classical particle. Equation (\ref{dotQ})
shows that $Q$ is a pure gauge variable: its dynamics remains
ambiguous since $N(t)$ is arbitrary. Like in the previous example,
one could fix the gauge by choosing $\dot{Q} = 1$, or any other
admissible gauge condition.

\bigskip

The procedure described in this subsection can be used to
parametrize the Lagrangian (\ref{newlagrangian}). In such case, the
so built parametrized Lagrangian will have $3N+1$ canonical
variables, since $Q$ is added to the $N$ vectors $\mathbf{r}_k$,
together with $7$ first class constraints. The number of degrees of
freedom will then result $3N-6$. In this sense, any system
displaying an absolute time, like the one described by
Lagrangian(\ref{newlagrangian}), could be regarded as a gauge-fixed
parametrized system.

\subsection{Jacobi action}

A different approach to the problem of eliminating the absolute time
consists in regarding the clock as a piece of the original
mechanical system. In such case, we will not add a canonical
variable $Q$ to the system of $N$ particles. Instead, we will try to
relate the evolution of the rest of the system to that piece of the
system playing the role of clock. If convenient, a parameter $t$
will be still introduced; but the action cannot be sensitive to it.
The action should contain only positions and velocities of the
particles and, at the same time, behave as a parametrized system.
These features are fulfilled by the Jacobi-like action
\cite{Lanczos} whose Lagrangian is
\cite{Barbour75,Barbour77,Barbour82}
\begin{equation}
L~=~2\,\sqrt{\Lambda -V}\,\sqrt{T}~,  \label{Jacobi}
\end{equation}%
where $\Lambda $ is a constant and $T$ is the \textit{compensated} kinetic
energy of Lagrangian (\ref{newlagrangian}). In fact, $T$ is made up of terms
quadratic in the velocities; then, $\sqrt{T}\ dt$ is invariant under
reparametrizations. In this case the canonical momenta are%
\begin{equation}
\mathbf{p}_{k}\ =\ \frac{\sqrt{\Lambda -V}}{\sqrt{T}}\ \frac{\partial T}{%
\partial \mathbf{v}_{k}}~,
\end{equation}%
where $\partial T/\partial \mathbf{v}_{k}$ are the momenta obtained in Eq.~(%
\ref{newmomentum}); they satisfy (see Eq.~(\ref{newlagrangian2}))
\begin{equation}
\sum_{k}\frac{1}{2\,m_{k}}\ \frac{\partial T}{\partial \mathbf{v}_{k}}\cdot
\frac{\partial T}{\partial \mathbf{v}_{k}}\ =\ T~.
\end{equation}%
Therefore we obtain the Hamiltonian constraint
\begin{equation}
\mathcal{H}\ =\ \sum_{k}\,\frac{\mathbf{p}_{k}\cdot \mathbf{p}_{k}}{2\ m_{k}}%
+V-\Lambda \ =\ 0~.  \label{Lambda}
\end{equation}%
Thus $\Lambda $ looks as the total energy of the universe. However,
differing from Jacobi's idea, its value cannot be chosen within the
initial conditions because $\Lambda $ enters the action as a
universal constant. Furthermore, the initial conditions are
constrained to fulfill the Eq.~(\ref{Lambda}) for a given value of
$\Lambda$. Instead, the constraint (\ref{classicalH}) allowed the
total energy to take any value because the initial value of $P$ was
not constrained. This is the only difference between both
approaches. So this approach displays one degree of freedom less
than the former: since the number of variables has not been
increased, the number of degrees of freedom reduces to $3N-7$.
Otherwise the dynamics is still governed by the equations
(\ref{HE}), because the form of $\mathcal{H}$ as a function of
$\mathbf{r}_{k}$, $\mathbf{p}_{k}$ is the same in both cases. It can
be verified that Lagrange equations recover their usual form too,
since the constraint implies that the factor $\sqrt{\Lambda
-V}/\sqrt{T}$ is equal to $1$. Concerning the choice of an internal
time $\tau (\mathbf{r}_{k},\mathbf{p}_{k})$, see Footnote
\ref{note}.

\section{Scale invariance}

\label{scaleinvariance}

In recent years much attention has been focused on \textit{shape-dynamics }%
\cite{Barbour03,Mercati,Anderson}, which is a kind of relational
mechanics with a local scale invariance. The scale transformation
is%
\begin{equation}
\mathbf{r}_{i}~\longrightarrow (1+\lambda )~\mathbf{r}_{i}~.  \label{scale}
\end{equation}%
This transformation is generated by%
\begin{equation}
\mathcal{G}\doteq \sum\limits_{k}~\mathbf{p}_{k}\cdot \mathbf{r}_{k}~,
\end{equation}%
since $\delta \mathbf{r}_{i}=\lambda
\{\mathbf{r}_{i~},\mathcal{G}\}$. In Ref.~\onlinecite{Barbour03}
$\mathcal{G}$ is called the \textit{dilatational momentum}. Scale
transformation should not be confused with a change of units. A
change of units not only scales positions and velocities but also
changes the fundamental constants entering the Lagrangian, in such a
way that the Lagrangian is affected only by a global harmless
factor. Instead, a (rigid) scaling only affects positions and
velocities; so one cannot expect an inoffensive effect. In general,
Lagrangians are not invariant under rigid
scalings. However, let us consider the parametrized Lagrangian (\ref{Jacobi}%
) in the case $\Lambda =0$. This Lagrangian would be invariant under
rigid scalings if the interaction potentials were inversely
proportional to the squared distances. In fact, under rigid scalings
the compensated kinetic
energy transforms as%
\begin{equation}
T~\longrightarrow (1+\lambda )^{2}~T
\end{equation}%
($\mathbf{v}_{k}\rightarrow (1+\lambda )\,\mathbf{v}_{k\ }$; $\mathbf{J}%
\cdot \mathbf{I}^{-1}\cdot \mathbf{J}$ is homogeneous of degree $2$ in the
velocities but homogeneous of degree $0$ in the positions). So the behavior
of $T$ under rigid scalings is compensated by the potential only if $%
V_{ij}=\kappa /r_{ij}^{2}$. We will now consider the gauging of the
scale invariance. \footnote{%
In the case of a conventional Lagrangian $L=T-V$, where $V$ were
inversely proportional to the squared distances, the local scaling
$L\rightarrow (1+\lambda
(t))^{2}~T-V/(1+\lambda (t))^{2}$ would count as a reparametrization $%
N(t)^{-1}=(1+\lambda (t))^{2}$.} Since $\mathcal{G}$ is going to be the
generator of a gauge transformation, we expect the constraint
\begin{equation}
\mathcal{G}=0.  \label{Gconstraint}
\end{equation}%
Concerning $\mathcal{P}_{\mu \,}$ and $\mathcal{J}_{\nu }$,
$\mathcal{G}$ is\ a first class constraint; in fact, $\mathcal{G}$
is invariant under rotations and $\{\mathcal{G},\mathcal{P}_{\mu
\,}\}=\mathcal{P}_{\mu \,}$. Besides $\{\mathcal{G},\mathcal{H}\}$
vanishes on the constraint surface in the considered case. In fact
\begin{equation}
\{\mathcal{G},\mathcal{H}\}~=~\{\sum\limits_{k}\,\mathbf{p}_{k}\cdot \mathbf{%
r}_{k}~,\sum\limits_{i}\,\frac{\mathbf{p}_{i}\cdot \mathbf{p}_{i}}{2~m_{i}}%
+V\}~=~\sum\limits_{k}\,\left( \frac{\mathbf{p}_{k}\cdot \mathbf{p}_{k}}{%
m_{k}}-\mathbf{r}_{k}\cdot \mathbf{\nabla }_{k}V\right) ~=~2\ (\mathcal{H}%
-V)-\sum\limits_{k}\,\mathbf{r}_{k}\cdot \mathbf{\nabla }_{k}V~,
\end{equation}%
where%
\begin{eqnarray}
\sum\limits_{k}\,\mathbf{r}_{k}\cdot \mathbf{\nabla }_{k}V~
&=&~\sum\limits_{k}\sum\limits_{i<j}\,\mathbf{r}_{k}\cdot \mathbf{\nabla }%
_{k}V_{ij}(r_{ij})~=~\frac{1}{2}\sum\limits_{k}\sum\limits_{i\neq j}\,\frac{%
\partial V_{ij}}{\partial r_{ij}}~\mathbf{r}_{k}\cdot \mathbf{\nabla }%
_{k}r_{ij}  \notag \\
&=&~\frac{1}{2}\left( \sum\limits_{k\neq j}\,\frac{\partial V_{kj}}{\partial
r_{kj}}~\frac{\mathbf{r}_{k}\cdot \mathbf{r}_{kj}}{r_{kj}}~\mathbf{-}%
\sum\limits_{k\neq i}\,\frac{\partial V_{ik}}{\partial r_{ik}}~\frac{\mathbf{%
r}_{k}\cdot \mathbf{r}_{ik}}{r_{ik}}\right) ~=~\frac{1}{2}\sum\limits_{k\neq
j}\,\frac{\partial V_{kj}}{\partial r_{kj}}~\frac{\mathbf{r}_{kj}\cdot
\mathbf{r}_{kj}}{r_{kj}}~=~\sum\limits_{k<j}\,\frac{\partial V_{kj}}{%
\partial r_{kj}}~r_{kj}~.
\end{eqnarray}%
Since $V_{kj}\propto r_{kj}^{-2}$ one obtains
\begin{equation}
\sum\limits_{k}\,\mathbf{r}_{k}\cdot \mathbf{\nabla }_{k}V~=~-2 ~V.
\end{equation}%
Thus one gets the vanishing of $\{\mathcal{G},\mathcal{H}\}$ on the
constraint surface: \footnote{\label{note}Instead the Newtonian
potential $V_{kj}\propto r_{kj}^{-1}$ leads to
$\{\mathcal{G},\mathcal{H}\}=2 \mathcal{H}-V$. In such case,
$\mathcal{G}$ monotonically increases on-shell ($\mathcal{H}=0$)
since the Newtonian potential is negative. Because of this reason,
$\tau=\mathcal{G}$ is a good internal time for a classical
self-gravitating system. This property has been exploited in
Ref.~\onlinecite{Barbour14} to handle the \textit{problem of time}
in quantum gravity \cite{Kuchar} within the framework of a
dimensionless formulation.}
\begin{equation}
\{\mathcal{G},\mathcal{H}\}~=~2\ \mathcal{H}\,~.
\end{equation}

Let us study the compensating term we should add to the Lagrangian (\ref%
{newlagrangian}) for gauging the scale invariance. Since velocities change as%
\begin{equation}
\mathbf{v}_{k}~\longrightarrow (1+\lambda )~\mathbf{v}_{k}\,+\,\dot{\lambda}~%
\mathbf{r}_{k}~,
\end{equation}%
the infinitesimal change of the intrinsic kinetic energy is%
\begin{equation}
\sum\limits_{i<j}\,\frac{m_{i}~m_{j}}{M}~\mathbf{v}_{ij}\cdot \delta \mathbf{%
v}_{ij}~=~\lambda \,\sum\limits_{i<j}\,\frac{m_{i}~m_{j}}{M}~\mathbf{v}%
_{ij}\cdot \mathbf{v}_{ij}\,+\,\dot{\lambda}\sum\limits_{i<j}\,\frac{%
m_{i}~m_{j}}{M}~\mathbf{v}_{ij}\cdot \mathbf{r}_{ij}~.
\end{equation}%
This behavior can be compensated by the term $I^{-1}G^{2}/2$ , where $G$ is
the \textit{intrinsic virial }(the virial in a center-of-mass frame):%
\begin{equation}
G~\doteq ~\sum\limits_{i<j}\,\frac{m_{i}~m_{j}}{M}~\mathbf{v}_{ij}\cdot
\mathbf{r}_{ij}~=~\sum\limits_{k}\,m_{k}~\mathbf{v}_{k}\cdot \mathbf{r}%
_{k}\,-\,\mathbf{P}\cdot \mathbf{R}~=~\sum\limits_{k}\,m_{k}~\mathbf{v}%
_{k}\cdot (\mathbf{r}_{k}-\mathbf{R)~},
\end{equation}%
and $I$ is the intrinsic\textit{\ scalar }moment of inertia:%
\begin{equation}
I~\doteq ~\sum\limits_{i<j}\,\frac{m_{i}~m_{j}}{M}~r_{ij}^{2}~~=~\sum%
\limits_{k}\,m_{k}~r_{k}^{2}\,-\,MR^{2}~~=~\sum\limits_{k}\,m_{k}~|\mathbf{r}%
_{k}-\mathbf{R}|^{2}~.
\end{equation}%
Both $G$ and $I$ are invariant under local rotations and translations; so,
they do not interfere the local invariance of the kinetic energy. In $%
I^{-1}G^{2}$ only the behaviors of the velocities are relevant under
scaling, because the scaling factors coming from the $\mathbf{r}_{ij}$'s
compensate each other. Therefore%
\begin{equation}
\delta \left( I^{-1}G^{2}/2\right) ~=~I^{-1}G~\delta
G~=~I^{-1}G~\sum\limits_{i<j}\,\frac{m_{i}~m_{j}}{M}~(\lambda ~\mathbf{v}%
_{ij}+\dot{\lambda}~\mathbf{r}_{ij}~)\cdot \mathbf{r}_{ij}~=~\lambda
~I^{-1}G^{2}\,+\,\dot{\lambda}~G
\end{equation}%
Besides the term $-(1/2)~\mathbf{J}\cdot \mathbf{I}^{-1}\cdot \mathbf{J}$ in
(\ref{newlagrangian}) changes because the velocities change (the scaling
factors coming from $\mathbf{r}_{ij}$'s compensate each other). So, we just
consider the change%
\begin{equation}
\delta \mathbf{J~}=~\sum\limits_{i<j}\,\frac{m_{i}~m_{j}}{M}~\mathbf{r}%
_{ij}\times (\lambda ~\mathbf{v}_{ij}+\dot{\lambda}~\mathbf{r}%
_{ij}~)~=~\lambda ~\mathbf{J~,}
\end{equation}%
which coincides with a rigid change;\ so it does not need a compensating
term. Thus the kinetic energy%
\begin{equation}
T~~=~\sum\limits_{i<j}\,\frac{m_{i}~m_{j}}{2M}~\mathbf{v}_{ij}\cdot \mathbf{v%
}_{ij}~-\,\frac{1}{2}~\mathbf{J}\cdot \mathbf{I}^{-1}\cdot \mathbf{J}\,-\,%
\frac{1}{2}~I^{-1}~G^{2}~  \label{Tscaling}
\end{equation}%
behaves under local scalings just as under rigid scalings. The conjugated
momenta
\begin{equation}
\mathbf{p}_{k}\ =\ m_{k}\left[ \mathbf{v}_{k}-\frac{\mathbf{P}}{M}-(\mathbf{I%
}^{-1}\cdot \mathbf{J})\times (\mathbf{r}_{k}-\mathbf{R})-I^{-1}~G~(\mathbf{r%
}_{k}-\mathbf{R})\right] ~  \label{newnewmomentum}
\end{equation}%
satisfy, as expected, the constraint (\ref{Gconstraint}). Moreover, the new
compensating term $I^{-1}~G~(\mathbf{r}_{k}-\mathbf{R})$ does not modify the
constraints (\ref{constraints}). The momenta (\ref{newnewmomentum}) contain
the particle velocities relative to a frame based at the center of mass that
rotates with the mean angular velocity of the system, diminished by a sort
of mean radial velocity of the system. Therefore, the global expansion has
been deducted from the particle velocities. As a consequence, the momenta
have the same behavior under rigid and local scalings, namely $\delta
\mathbf{p}_{k}=~\lambda \mathbf{\ \{p}_{k\,},\mathcal{G}\}$. Kinetic energy (%
\ref{Tscaling}) can be rewritten as \footnote{%
In Eq.~(\ref{newnewmomentum}), the terms entering the bracket to compensate
rotations and expansions have the general structure $-I^{-1}A\,B_{k\ }$,
such that $\sum m_{k}\,v_{k}\,B_{k\ }=A$ and $\sum m_{k}\,B_{k\ }^{2}=I$.
Besides, both compensating terms are mutually orthogonal. These are the
essential ingredients to obtain just contributions $-(1/2)\,I^{-1}A^{2}$
when performing the square of $\mathbf{p}_{k\,}$. The somewhat different
structure of the compensating term $\mathbf{P}/M$ is due to the fact that we
started from an action that is already invariant under Galileo
transformations, which is a particular case of local translations.}%
\begin{equation}
T~~=~\sum\limits_{k}\,\frac{m_{k}}{2}~\left\vert \mathbf{v}_{k}-\frac{%
\mathbf{P}}{M}-(\mathbf{I}^{-1}\cdot \mathbf{J})\times (\mathbf{r}_{k}-%
\mathbf{R})-I^{-1}~G~(\mathbf{r}_{k}-\mathbf{R})\right\vert ^{2}~.
\end{equation}%
The system has $3N$ canonical coordinates and $8$ first class constraints\ $%
\mathcal{P_{\mu \,}}$,$~\mathcal{J_{\nu \,}}$, $\mathcal{G}$, $\mathcal{H}$.
So, $3N-8$ genuine degrees of freedom are left in shape-dynamics. For
instance, $3$ particles would have $2$ degrees of freedom in shape-dynamics,
that relate to the two angles one needs to define a triangle regardless of
its size and orientation; but one of these degrees of freedom is related to
the internal time.

\bigskip

\section{Conclusions}

\label{conclusions}

Newton's mechanics relies on external symmetries linked to the
concepts of absolute space and time. These symmetries manifest
themselves through the idea of inertial frames together with the
Galilean group as a tool for relating coordinates belonging to
different inertial frames. Instead, Relational Mechanics describes
the system by means of equations that can be used in any frame. This
means that Relational Mechanics is not involved with absolute
magnitudes but only with the relative motions of the particles of
the system. This goal is reached by gauging the Galilean group;
i.e., by adding compensating terms that makes the kinetic energy
behave in the same way under rigid and local translations and
rotations. The compensating terms do not contain external fields but
are built with the own variables of the system, in such a way that
the gauge invariant Lagrangian results to be written just in terms
of intrinsic quantities (notice that intrinsic quantities
(\ref{kinetic}), (\ref{J}) and (\ref{Itensor}) look like the
respective usual quantities in a center-of-mass frame). As in any
gauge theory, the generators of gauge transformations --3
translations and 3 rotations in this case-- become first class
constraints. So, the system loses 6 degrees of freedom since global
time-dependent translations and rotations have been excluded from
the category of \textit{motions}: they do not constitute changes of
the relations between particles. In the Lagrange formulation of
Relational Mechanics, two sets of initial conditions that differ in
a gauge transformation (a change to an arbitrarily moving frame) are
equivalent; in particular, any configuration corresponding to a
rigid motion is equivalent to the state of rest. In the Hamiltonian
formulation, the losing of 6 degrees of freedom is evidenced in the
fact that the set of initial data for the canonical variables is
constrained by 6 constraint functions.

\bigskip

Differing from the absolute space, absolute time is not eliminated
by means of counterterms that covariantize the time derivative. The
gauging of time translations affects the operator $d/dt$ itself. So,
we have to resort to forms of Lagrangian that reproduce the usual
dynamics but leave the parameter $t$ as an irrelevant quantity that
can be arbitrarily changed; these are the parametrized Lagrangians.
In particular, the Jacobi action describes a parametrized system
which acquires scale invariance if the potentials are inversely
proportional to the squared distances. This rigid invariance can be
gauged by introducing a new counterterm in the kinetic energy. The
result is the Lagrangian for the so called shape-dynamics.

\bigskip

Relational equations of motion (\ref{motion}) exhibit two types of
interaction. On the one hand, the potential $V$ provides gauge
invariant forces between pairs of particles. On the other hand, each
particle interacts with the entire universe through gauge dependent
inertial forces. The unquestionable success of Newton's laws can be
fully understood in the framework of Relational Mechanics: Newton's
laws are valid in any frame where the center of mass of the universe
moves at a constant velocity, and the total intrinsic angular
momentum of the universe $\mathbf{J}$ vanishes (however, Newtonian
solutions must be filtered by the gauge condition $\mathbf{J}=0$).
This result realizes Mach's principle, in the sense that these
privileged Newtonian frames are selected by the distribution of
matter in the universe. The origin of the inertial forces lies in
the entire universe, as seen in the example of Newton's bucket.

\bigskip

In his famous article of 1916, Einstein began \S 2 by saying
\textit{``In classical mechanics, and no less in the special theory
of relativity, there is an inherent epistemological defect which
was, perhaps for the first time, clearly pointed out by Ernst
Mach''} \cite{Einstein16}. He then paraphrased Newton's thought
experiment of the rotating bucket by proposing two equally
constituted fluid bodies in relative rotation around the axis
joining their centers, at a relative distance enough large to ignore
the gravitational interaction. Einstein said that if one of the
bodies resulted to be spherical and the other one an ellipsoid of
revolution, then it should exist a physical reason other than the
rotation with respect to the absolute space for such a difference.
Relational Mechanics can tackle the issue because no difference is
expected if the bodies are the sole constituents of the universe.
Instead, the rotation relative to the rest of the universe will
produce the equatorial deformation.

\begin{acknowledgments}
This work was supported by Consejo Nacional de Investigaciones Cient\'{\i}ficas y
T\'{e}cnicas (CONICET) and Universidad de Buenos Aires.
\end{acknowledgments}

\end{document}